\documentclass[12pt]{article}
\usepackage{amsmath,amsthm,amsfonts}
\usepackage{scicite}
\usepackage{times}
\usepackage{color}
\usepackage{comment}
\usepackage{bm}
\usepackage{physics}

\usepackage{float}
\usepackage{graphicx}
\usepackage{afterpage}
\usepackage{caption}
\usepackage{subcaption}
\newenvironment{sciabstract}{
\begin{quote} \bf}
{\end{quote}}

\usepackage{subfiles}
\usepackage{titling}

\theoremstyle{definition}
\usepackage{tikz}
\definecolor{sing}{RGB}{46, 134, 171}
\definecolor{trip}{RGB}{162, 59, 114}
\definecolor{quin}{RGB}{241, 143, 1}
\definecolor{sept}{RGB}{199, 62, 29}
\definecolor{none}{RGB}{59, 31, 43}

\usepackage[font=small,labelsep=quad,format=plain]{caption}
\usepackage{hyperref}
\hypersetup{colorlinks=true,citecolor=blue,linkcolor=blue,urlcolor=blue}

\renewcommand{\figurename}{Fig.}

\begin{document}

\title{Simulating challenging correlated molecules and materials on the Sycamore quantum processor}

\author
{Ruslan N.~Tazhigulov,$^{1}$ Shi-Ning Sun,$^{2}$ Reza Haghshenas$^{1}$, Huanchen Zhai,$^{1}$ \\
Adrian~T.~K.~Tan,$^{2}$ Nicholas C.~Rubin,$^{3}$ Ryan Babbush,$^{3}$ Austin~J.~Minnich,$^{2}$\\
Garnet~Kin-Lic Chan$^{1}$\\
{\normalsize $^{1}$Division of Chemistry and Chemical Engineering, California Institute}\\ \normalsize{of Technology, Pasadena, California 91125, USA}\\
\normalsize{$^{2}$Division of Engineering and Applied Science, California Institute of}\\ \normalsize{Technology, Pasadena, California 91125, USA}\\
\normalsize{$^{3}$Google Quantum AI, 340 Main Street, Venice, California 90291, USA}
}
\date{}

\baselineskip24pt

\maketitle

\begin{sciabstract}
Simulating  complex molecules  and  materials  is  an anticipated
application  of  quantum  devices.  With strong quantum advantage demonstrated in artificial tasks, we examine how such advantage translates into modeling physical problems of correlated electronic structure. 
We simulate static and dynamical electronic structure on a  superconducting  quantum  processor  derived from Google’s Sycamore architecture for two representative correlated electron problems: the nitrogenase iron-sulfur molecular clusters, and $\alpha$-ruthenium trichloride, a proximate spin-liquid material. To do so, we simplify the electronic structure into low-energy spin models that fit on the device. 
With extensive error mitigation and assistance from classically simulated data, we achieve quantitatively meaningful results deploying about 1/5 of the gate resources used in artificial quantum advantage experiments on a similar architecture. This increases to over 1/2 of the gate resources when choosing a model that suits the hardware. Our work serves to convert artificial measures of quantum advantage into a physically relevant setting.
\end{sciabstract}

\section*{Introduction}

There has been much interest in simulating problems of chemistry and materials science on quantum computers \cite{Chan_ChemRev_2020,Yuan_RevModPhys_2020}.
This is not least because the first demonstrations of quantum ``supremacy'' in a number of tasks have appeared \cite{Quantum_Supremacy_Google_Nature_2019,wu2021strong,zhong2020quantum}. 
Such simulations have deployed an impressive level of quantum resources;  the random circuit supremacy experiments on the Sycamore processor used more than 50 qubits, and approximately 500 two-qubit gates.
However, the common feature of current demonstrations of supremacy is their artificial nature. In particular, the achievement of supremacy is in part due to the selection of tasks and metrics designed to facilitate large quantum advantage on current hardware. For example, Ref.~\cite{Quantum_Supremacy_Google_Nature_2019} involved distinguishing a theoretical signal of 1 from 0, but a noisy experimental result of 0.002 was already sufficient to claim success due to the comparative classical complexity to obtain the same result. How transferable success in such a  metric is to generally accepted criteria for successfully simulating actual chemical and materials problems is unclear.

In this work, we study the simulation of two ``realistic'' (if still highly simplified) problems of strongly correlated molecular and materials electronic structure on a state-of-the-art superconducting quantum processor. 
Aside from the necessary simplifications, neither the systems nor the observables studied were chosen to favor the experimental hardware. The current work thus attempts to report on the ability of current quantum devices to tackle problems of real-world interest without careful preselection. In particular, we examine the relation between the deployable quantum resources to achieve artificial supremacy metrics and those that can be used in successful electronic structure simulations.

The first problem we consider is the low-energy electronic structure of the iron-sulfur clusters of nitrogenase, including the FeMo-cofactor, critical components of the natural nitrogen cycle. The second is the electronic structure of $\alpha$-RuCl$_3$, a candidate material for realizing spin-liquid physics. Both are strongly correlated electron structure problems that contain challenging features for quantitative heuristic quantum chemistry approximations. 
Although for practical reasons we need to reduce the electronic structure into low-energy spin models, the range, nature, and topology of the interactions provide some elements of real-world complexity that can be absent from more specially chosen  simulation problems. We target finite-temperature static and dynamical properties representative of accessible physical chemistry measurements, and study both systems using the Weber superconducting qubit processor derived from Google's Sycamore architecture, using the finite-temperature version of quantum imaginary time evolution with recompilation \cite{Motta_QITE_NatPhys2020, IBM_PRXQuantum2021}. To obtain meaningful data, many types of error mitigation are necessary, thus we discuss the impact of different protocols. Our results illustrate to what extent current superconducting quantum processors can be used to simulate real-world  chemical and materials problems.

\section*{Results}

\subsection*{Overview of systems}
\textbf{Fe-S clusters and nitrogenase.} Nitrogenases are enzymes that convert atmospheric dinitrogen into ammonia. The process involves the coordinated transfer of multiple electrons and protons to dinitrogen, and utilizes multiple metalloclusters found in the nitrogenase enzyme: the [4Fe-4S] Fe-cluster, the [8Fe-7S] P-cluster, and the [7Fe-1Mo-9S] FeMo-cofactor (the latter  can also be found with other metals replacing molybdenum)~\cite{beinert1997iron}. The electronic structure of the Fe-S clusters remains incompletely understood;  theoretical calculations on the Fe- and P-clusters unveiled a large number of low-lying spin-states~\cite{sharma2014low,li2019electronic}, whose role in the electron transfer process is unclear. In the case of the FeMo-cofactor, the oxidation and spin states of the ions are unresolved. 

\begin{figure*}[htbp]
\centering{
\includegraphics[width=0.95\textwidth]{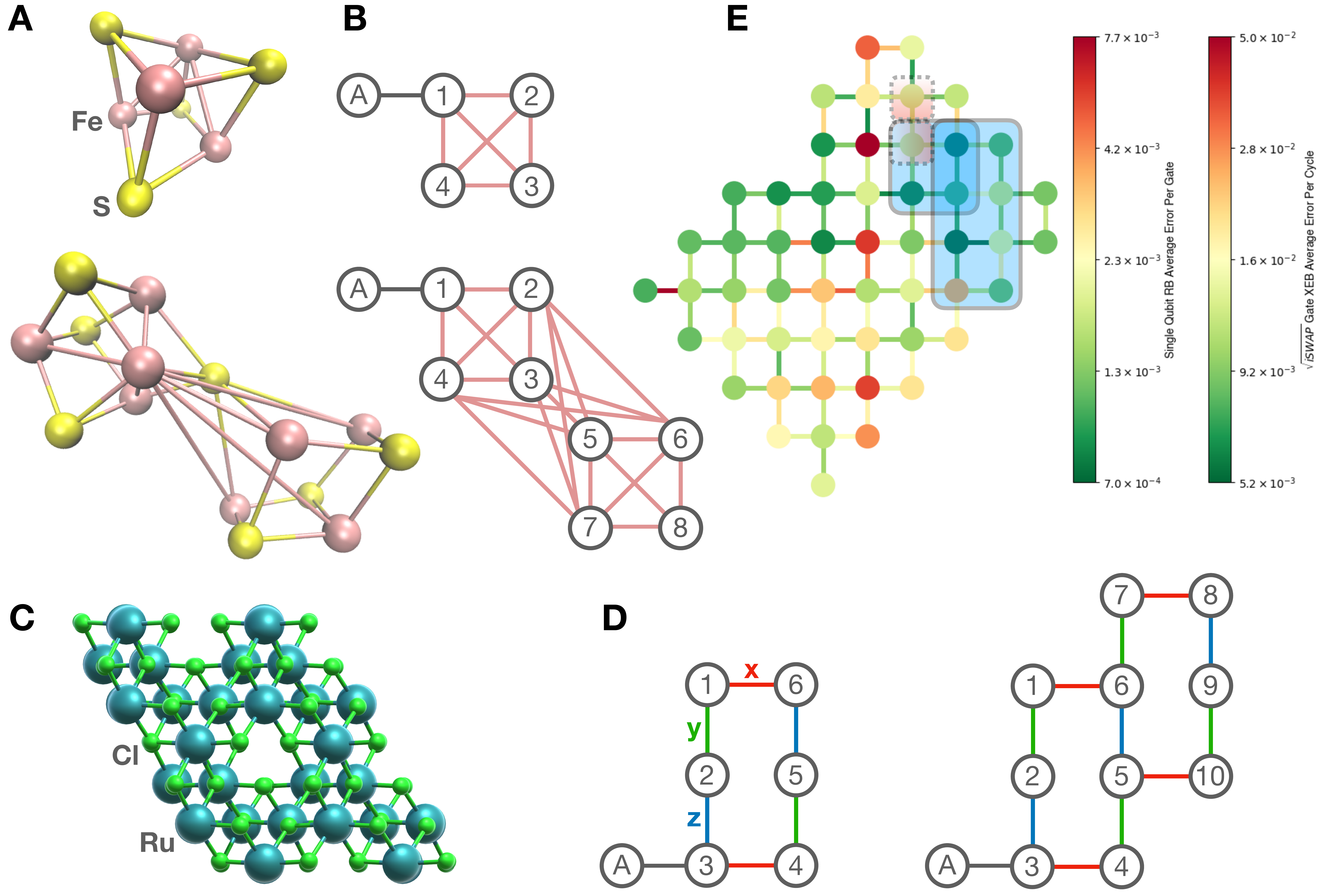}}
\caption{(A) Molecular structures of [4Fe-4S] and [8Fe-7S] P-cluster. (B) Topology of 4-site and 8-site Heisenberg spin models of [4Fe-4S] (top) and [8Fe-7S] P-cluster (bottom). (C) Crystal structure of $\alpha$-RuCl$_3$: trigonal space group P3$_1$12. (D) Topology of 6-site and 10-site Kitaev-Heisenberg spin models of $\alpha$-RuCl$_3$. The adjacent ancilla qubits are indicated as ``A". (E) Illustrative examples of qubit patches (blue) used for simulations of 4-site and 8-site models. The ancilla qubits (dotted boxes) are shown in red. The color scheme of qubits and connectivities represent ``single-qubit randomized benchmarking (RB) average error per gate'' and ``two-qubit $\sqrt{\text{iSWAP}}$ gate cross-enthropy benchmarking (XEB) average error per cycle'', respectively.}
\label{fig:models}
\end{figure*}

The simplest electronic models of these compounds involve the Fe and S valence orbitals, with over 100 spin-orbitals in the case of FeMo-co, too large for current quantum devices~\cite{li2019electronic2}. However, both theoretical and spectroscopic studies indicate that the electron-hole-like excitations mostly lie at higher energies than a manifold of low-energy spin-coupled states. Thus the low-lying electronic spectrum can be captured using a variety of spin models, which are commonly used to interpret the electronic structure~\cite{sharma2014low,noodleman1995orbital,yamaguchi1989antiferromagnetic} as well as to model measured quantities, such as the magnetic susceptibility and heat capacity~\cite{aizman1982electronic,chen2022using}. In the most commonly studied Fe-S models, the metals occupy vertices and the S bridges form the edges, with the latter enforcing antiferromagnetic interactions parametrized by the exchange coupling $J$~\cite{sharma2014low,noodleman1995orbital}. The 3D models and the 2D pattern of spin couplings are shown in Fig.~\ref{fig:models}. 
Although the metal ions have large total spin (e.g. $S=2, S=5/2$), we can further simplify (reducing the quantum resources) by representing each metal by a $S=1/2$ spin, retaining only the topology of the spin-spin couplings, and with an isotropic Heisenberg exchange interaction between each spin pair. Although the spectrum is changed by this reduction, it retains similar features in the lowest few states, e.g. the total spin value and degeneracy, as discussed in the supplementary information.
The simplified Hamiltonian can thus be written as 
\begin{equation}
\hat{H} = \sum_{ij} J_{ij} {S}_i \cdot {S}_j
\label{eq:heisLR}
\end{equation}
where $S_i$, $S_j$ are $S=1/2$ operators, and $\langle ij\rangle$ denotes each of the bonds in Fig.~\ref{fig:models}. We note that other instances of Heisenberg models have been simulated on quantum devices (see e.g.~\cite{chiesa2019quantum,kandala2017hardware}) but the interesting properties of the FeS clusters relate to the specific topology and magnitudes of the spin couplings.

In the case of [4Fe-4S] we used only two different $J_{ij}$, namely $J=1, J'=1.17J$ characteristic of the all-ferric cubane~\cite{moula2018synthesis} ($J'$ couples (1-2), (3-4), $J$ couples the other pairs, as shown in Fig.~\ref{fig:models}B), while in the P-cluster and FeMo-co, we used a single $J$ for all couplings (all $J$ positive). With this simplification, the P-cluster and FeMo-co have the same spin-Hamiltonian. (Note that the units chosen for energy define an inverse energy unit for time; all times will hence be assumed to be in such units).
 
\noindent \textbf{Ruthenium trichloride, $\alpha$-RuCl$_3$.} Ruthenium trichloride is a transition metal material of interest as a ``proximate'' spin-liquid~\cite{banerjee2016proximate}. Similarly to in the Fe-S clusters, the low-energy excitations are spin-excitations. In particular, the edge-sharing octahedral coordination around the Ru(III) ions (see Fig.~\ref{fig:models}), together with the spin-orbit coupling, leads to a low-energy Hamiltonian that approximates the exactly solvable Kitaev model in the spin-liquid regime. However, the degree of similarity to the Kitaev model, as well as the interpretation of spectroscopic and thermal measurements in terms of modifications to the Kitaev model, is much debated. 
 
Much work has been devoted to deriving and parametrizing low energy Hamiltonians for $\alpha$-RuCl$_3$~\cite{maksimov2020rethinking}. 
The simplest family of Hamiltonians are the Kitaev-Heisenberg models, that take the form
\begin{equation}
    \hat{H}_\text{KH} = J \sum_{\langle i,j\rangle} S_i S_j + K_\gamma \sum_{\langle i,j\rangle^\gamma} S^\gamma_i S^\gamma_j
    \label{eq:KH}
\end{equation}
where $\gamma$ = X, Y, Z.
We also use a parametrization in the literature from Refs.~\cite{Suga_KHaRucl3_PRB2018,Suga_KHaRucl3_PRB2018_Erratum,maksimov2020rethinking}, where the above form is augmented by additional couplings
\begin{align}
\begin{split}
    \hat{H}_\text{KH} &= J \sum_{\langle i,j\rangle} S_i S_j + K_\gamma \sum_{\langle i,j\rangle^\gamma} S^\gamma_i S^\gamma_j \\
    &\quad + \Gamma \sum_{\langle i,j\rangle} (S^\alpha_i S^\beta_j + S^\beta_i S^\alpha_j) \\
    &\quad + \Gamma' \sum_{\langle i,j\rangle} \sum_{\alpha \neq \gamma} (S^\gamma_i S^\alpha_j +  S^\alpha_i S^\gamma_j)
    \label{eq:KH_gamma_terms}
\end{split}
\end{align}
where $\gamma$ ($\alpha$, $\beta$) = X, Y, Z, and $\alpha$, $\beta$ take indices different from $\gamma$. 
Specifically, we use the parameters: $J$ = $-$1.53, $K=K_\gamma$ = $-$24.4, $\Gamma$ = 5.25, $\Gamma'$ = $-$0.95. The Kitaev point, which is exactly solvable, correspond to taking only the $K_\gamma$ term in the above models.
It is argued that in the parameter regime of $\alpha$-RuCl$_3$, both the excitations and heat capacity show echoes of the two kinds of Majorana fermions that exist at the exactly solvable point~\cite{gohlke2017dynamics,laurell2020dynamical}.

\subsection*{Implementation}

\noindent \textbf{Formal algorithm}. 
We simulate the finite-temperature energies and dynamical correlation functions of the systems.
The quantum imaginary time evolution (QITE) algorithm is used to prepare a sample of the finite-temperature state $\rho = e^{-\beta H}/\text{Tr}(e^{-\beta H})$. The basic idea in QITE is to prepare normalized imaginary time-evolved states $e^{-{\beta H}/2}|\Psi_i(0)\rangle = c_i(\beta) \cdot U_i (\beta/2)|\Psi_i(0)\rangle$ on the quantum device.
Here $c_i(\beta)$ is a normalization constant tracked classically, $U_i(\beta/2)$ is the unitary determined by the QITE procedure, and $|\Psi_i(0)\rangle$ is the initial state (here the $i$th  computational basis state). 
We are interested in the finite-temperature static and dynamical observables $\langle A\rangle_\beta = \mathrm{Tr} [\rho A$], $\langle A(t)B\rangle = \mathrm{Tr} [\rho A(t) B]$, computed as
\begin{equation}
    \langle \hat{A} \rangle_\beta = \frac{\sum_{i} P_i A_i}{\sum_{i} P_i}
    \label{eq:trace_eval_approx}
\end{equation}
where $P_i$ = $|c_i(\beta)|^2$
and $A_i$ = $\langle{\Psi_i (\frac{\beta}{2})} |\hat{A}| {\Psi_i (\frac{\beta}{2})}\rangle$. The expectation values $A_i$ are obtained by reading out Pauli operators (static observables) or from the Hadamard test (dynamic observables).

In the original QITE procedure, the unitaries are determined at each time step from a set of linear equations constructed from measurements on the quantum device. Similarly, the normalization weights $P_i$ are accumulated at each time step.
Once the imaginary time states $|\Psi_i(\frac{\beta}{2})\rangle$ are prepared, they can then be  propagated in real-time to yield states $|\Psi_i(\frac{\beta}{2}+t)\rangle$, with the real-time propagation unitary $U(t)$ generated e.g. by a Trotter evolution of the Hamiltonian. The schematic circuit for a dynamical simulation is shown in Fig.~\ref{fig:models}A.
In the current experiments, however, we use classical recompilation~\cite{IBM_PRXQuantum2021} to generate both the circuits $U(\beta/2)$ and $U(t)$ and the weights $P_i$. The details are described further below. 

\noindent \textbf{Implementation and error mitigation}. 

\noindent \emph{Hardware and qubit selection}. The simulations were run on Google's 53-qubit Weber processor based on the Sycamore architecture~\cite{Quantum_Supremacy_Google_Nature_2019}. The Fe-S cluster simulations used up to 5 and 9 qubits respectively, while $\alpha$-RuCl$_3$ used up to 7 and 11 qubits respectively. (In the above qubit counts, we include the one ancilla qubit used for the Hadamard test circuit). The best performing qubits were selected using a combination of single-qubit randomized benchmarking, two-qubit cross-entropy benchmarking, and Loschmidt echo metrics on the hardware; an example of the embedding of the qubits onto the Weber architecture for the Fe-S clusters is shown in Fig.~\ref{fig:models}E.

\begin{figure*}[!htbp]
\centering{
\includegraphics[width=0.7\textwidth]{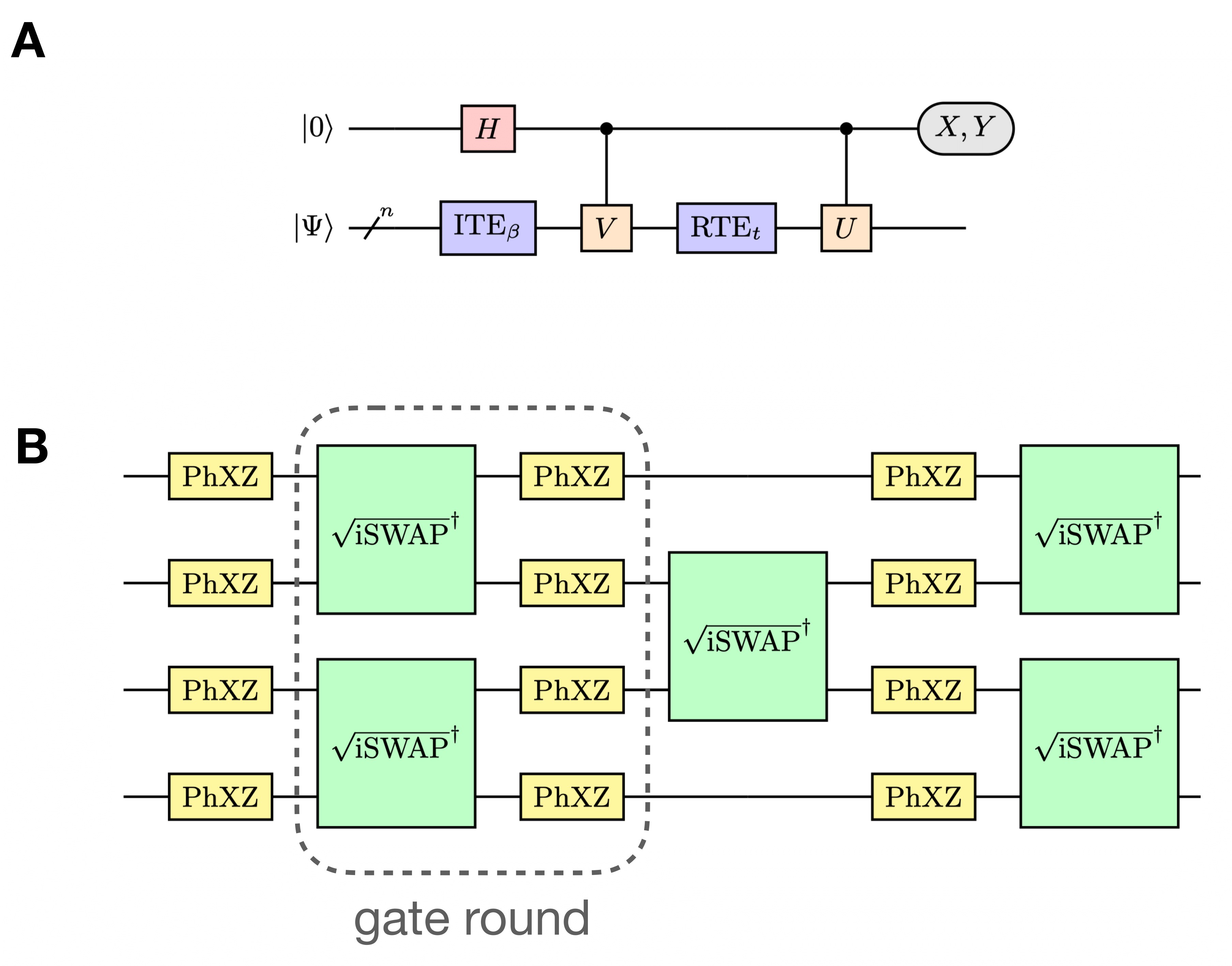}}
\caption{(A) Quantum circuit to calculate a finite-temperature dynamic correlation function $\langle \hat{U}(t) \hat{V} \rangle_\beta$ using QITE and one ancilla qubit. 
Measuring X and Y on the ancilla yields the real and imaginary parts of the correlation function, respectively. The thermal average over the initial states is obtained according to Eq.~\ref{eq:trace_eval_approx}. (B) Circuit recompilation ansatz. Each gate round after the base PhXZ layer includes a layer of two-qubit $\sqrt{\text{iSWAP}}^{\dagger}$ and single-qubit PhXZ gates as shown in the box. 
}
\label{fig:circuits_XY_XYZ_models}
\end{figure*}

\noindent \emph{Circuit recompilation}. The circuit realization
of the formal QITE algorithm (including $U(\beta/2)$ and $U(t)$) using a reasonable Trotter timestep (e.g. 0.1 inverse energy units) is far too deep for current quantum devices. For example, using a first-order Trotter expansion with a time step of 0.1 in the 6-site Kitaev-Heisenberg model at the Kitaev point, we already exceed the two-qubit gate count used in random circuit supremacy after 3 units of time. 
Thus, to reduce the circuit depth, for each desired imaginary time or imaginary plus real time point at which observables were to be measured, we defined the corresponding $U_i$ by classical recompilation. This is a variational procedure whereby the exact unitary is constructed classically, and then a circuit of given depth ($\bar{U}_i$) is (classically) variationally optimized to maximize the fidelity $||U_i|\Psi_i(0)\rangle - \bar{U}_i|\Psi_i(0)\rangle||$~\cite{IBM_PRXQuantum2021}. To represent $\bar{U}_i$ we chose a brickwork circuit of the native gates (single-qubit Phased XZ (PhXZ) and two-qubit $\sqrt{\text{iSWAP}}^{\dag}$) adjusted for systematic errors (see below), and we recompiled to a fidelity of $> 90\%$ in classical noiseless emulations (see Fig.~\ref{fig:circuits_XY_XYZ_models}B).

\noindent \emph{Post-selection}. The FeS clusters and simplified $\alpha$-RuCl$_3$ Hamiltonians possess $Z_2$ symmetry. We used this to perform postselection as discussed in Ref.~\cite{IBM_PRXQuantum2021}. 

\noindent \emph{Floquet calibration}. As discussed in Refs.~\cite{Google_FermiHubbard_arxiv_2020,Neill_quantumring_google_Nature_2021}, one can calibrate an
``excitation-number'' conserving gate in terms of 5 angles $\theta, \phi, \zeta, \gamma, \chi$. The ideal $\sqrt{\text{iSWAP}}^\dag$ gate should have $\theta = \pi / 4$, with all other angles being zero, but in practice this is not the case. We performed ``Floquet'' calibration~\cite{Google_FermiHubbard_arxiv_2020,Neill_quantumring_google_Nature_2021} before each experiment to calibrate the actual angles for each run. 
The modified $\sqrt{\text{iSWAP}}^\dag$ gate was then used in the classical recompilation procedure to obtain the most faithful compressed circuit. 

It is important to note that this calibration was only performed on isolated $\sqrt\text{iSWAP}^\dag$ gates. However, due to the specifics of the hardware, the presence of PhasedXZ gates (``microwave'' gates) can lead to additional errors in the two-qubit gates that are not accounted for in the calibration.

\noindent \emph{Dynamical decoupling}. In the schematic shown in Fig.~\ref{fig:circuits_XY_XYZ_models}A for dynamical observables, the ancilla qubit in the Hadamard test is idle for large parts and can decohere. To mitigate this effect, we employed a dynamical decoupling sequence~\cite{Suter_PRL2011_DD,Suter_PRL2014_DD} consisting of inserting identities generated by XX and YY gates on the ancilla. 

\noindent \textit{Rescaling}. To improve the data, we performed postprocessing, rescaling all dynamical correlation functions. To do so, we performed a classical noiseless simulation using a modified Hamiltonian $\hat{H}'$ containing only commuting terms on pairs of qubits. For example, for the hexagonal model shown in Fig.~\ref{fig:models}D, we used only the Hamiltonian terms on (1-2), (3-4), (5-6). The commuting form of $\hat{H}'$ means that the exact classical simulation can be performed easily.
We then performed the same simulation on the quantum device, generating the circuit by recompilation with the same ansatz as for the full Hamiltonian, and including all the error mitigation techniques described above. It is important to note that even though $\hat{H}'$ contains only commuting terms, the ansatz used for the quantum device did not have this simplifying commuting structure e.g. in Fig.~\ref{fig:models}B, the $\sqrt\text{iSWAP}^\dag$ gates couple all the qubits; thus the quantum simulation of $\hat{H}'$ samples similar noise to that of $\hat{H}$.  
For each time point $t$ (in imaginary and in real time), we then defined a rescaling factor
\begin{equation}
    f(t) = \frac{\langle A \rangle_t^\text{ideal}(\hat{H}')}{\langle A \rangle_t^{\text{hardware}}(\hat{H}')}
    \label{eq:rescaling}
\end{equation}
where $\langle A \rangle_t^\text{ideal}(\hat{H}')$ was obtained using classical noiseless emulation, and $\langle A \rangle_t^{hw}(\hat{H}')$ was obtained from the hardware. 
$f(t)$ was then used to rescale the dynamical correlation functions of the real Hamiltonian.

\subsection*{Simulations}

\begin{figure*}[!htbp]
\centering{
\includegraphics[width=0.95\textwidth]{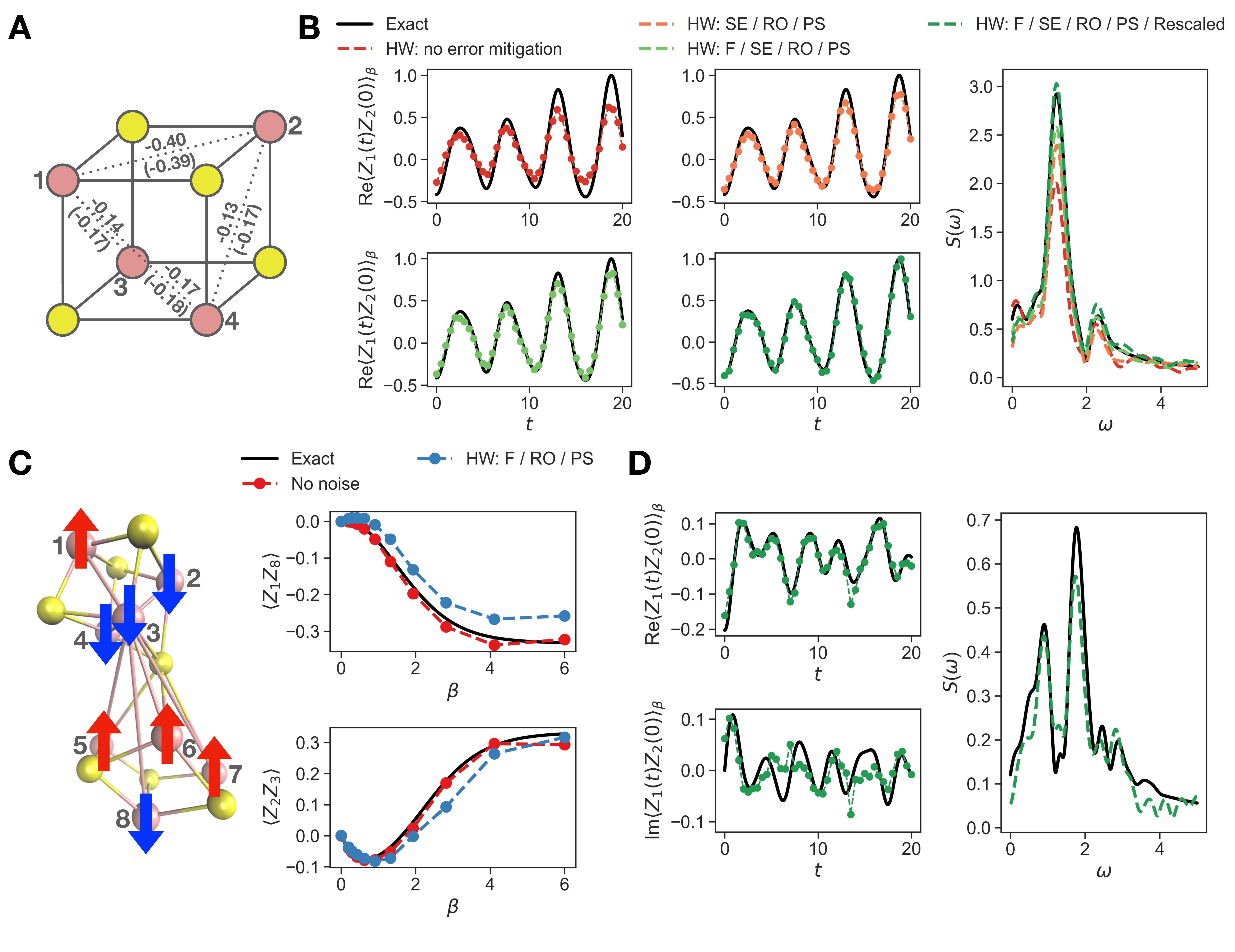}}
\caption{(A) Finite-temperature $\langle Z_i Z_j \rangle$ ($\beta$ = 2) correlation functions computed for the [4Fe-4S] model. The simulation values are in good agreement with the exact results in parentheses.
(B) Finite-temperature dynamical correlation function $\langle Z_1(t)Z_2(0) \rangle_\beta$ ($\beta$ = 2) and its Fourier transform for the [4Fe-4S] model. 
(C) Spin coupling pattern (derived from $\langle Z_i Z_j\rangle$ correlation functions) in the P-cluster/FeMo-co model, and simulated finite-temperature $\langle Z_1 Z_8 \rangle$ and $\langle Z_2 Z_3 \rangle$ for the model, showing the correct magnetic coupling pattern in the low temperature regime. The deviation between the exact and no-noise results show the effects of recompilation. (D) Real and imaginary parts of the finite-temperature dynamical correlation function $\langle Z_1(t)Z_2(0) \rangle_\beta$ ($\beta$ = 1) and its Fourier transform for the P-cluster/FeMo-co model.
(Error mitigation acronyms: spin echoes (SE), Floquet calibration (F), readout error mitigation (RO), post-selection (PS)).
}
\label{fig:fes}
\end{figure*}

\noindent \textbf{Fe-S clusters.} 
Fig.~\ref{fig:fes} shows simulation results on the Fe-S clusters. Despite the simplifications of the model (e.g. to $S=1/2$) the static $Z_iZ_j$ correlation functions at the lowest temperature ($\beta$ = 2) for the [4Fe-4S] and P-/FeMo-co clusters reflect the known ground-state pairing patterns of the spins in the true [4Fe-4S] and P-clusters (Figs.~\ref{fig:fes}A, C). The largest error in the correlation functions is approximately 22\% in the P-/FeMo-co cluster. The majority of this error is from the hardware rather than the classical recompilation; the error from classical recompilation can be seen in the difference between the red and black lines in Fig.~\ref{fig:fes}C.

Fig.~\ref{fig:fes}B shows the dynamical correlation functions and the effects of the different error mitigation strategies. The total circuit complexity for the [4Fe-4S] cluster dynamical correlation function is 22 two-qubit gates after circuit recompilation of the imaginary-time evolution and real-time evolution blocks and decomposition of the controlled gates (Fig.~\ref{fig:circuits_XY_XYZ_models}A) into $\sqrt{\text{iSWAP}}^\dagger$ gates; as discussed the recompilation provides nearly exact results in the classical noiseless emulator, indistinguishable from the exact results on the plot.
Even without error mitigation the dynamics displays the right frequencies (Fig.~\ref{fig:fes}B), but the amplitudes of the peaks are reduced. (We obtain some insight into the
hardware performance by modeling the unmitigated results using a depolarizing noise channel in the supplementary information).
The highest amplitude peak in the spectrum ($\omega$ = 1.17) represents a transition between the second excited state and ground state. 
Error mitigation reduces the error in the peak height by about 50\%, while postprocessing (rescaling) creates an almost perfect amplitude.

The total recompiled circuit complexity for the P-/FeMo-co cluster dynamical correlation functions is 82 two-qubit gates. 
The larger circuit complexity is reflected in the somewhat lower quality of the dynamical correlation functions. Up to frequency $\omega = 3$ (units of energy), there are 7 identifiable peaks in the spectrum. Encouragingly, the hardware results even before rescaling capture the correct frequency of most of these peaks, however, the amplitudes are severely degraded: the largest amplitude peak is only about 40\% of the expected height.
Post-processing the data (rescaling as well as shifting to satisfy  the imaginary part to zero mean) significantly improves the results, although the largest amplitude peak still has an amplitude error of 20\%.


\begin{figure*}[!htbp]
\centering{
\includegraphics[width=0.95\textwidth]{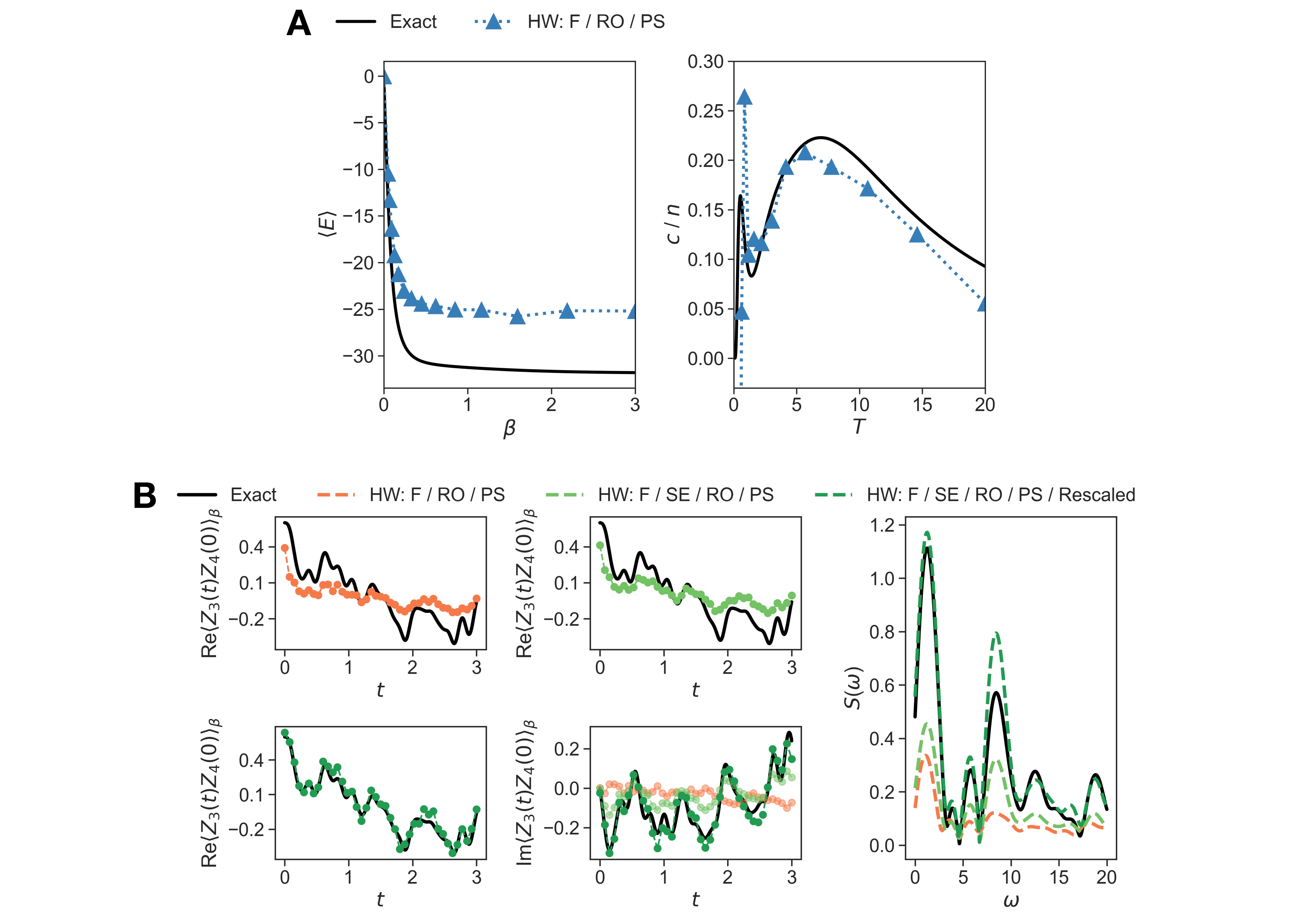}}
\caption{
(A) Thermal energy $\langle E \rangle$ and heat capacity ($c / n=\frac{1}{n}\partial \langle E\rangle/\partial T$) for the $\alpha$-RuCl$_3$, 6-site model ($n=6$).
The two peak structure is indicative of the proximate spin-liquid character~\cite{Loidl_PRB2019,Motome_PRL2014,Motome_PRB2015}.
(B) Dynamical correlation function $\langle Z_3(t)Z_4(0)\rangle_\beta$ and its transform $S(\omega)$ at $\beta$ = 1 for $\alpha$-RuCl$_3$, 6-site model. 
The 3 plots for the real part show the successive improvement obtained by standard error mitigation (Floquet, read-out error, post-selection, orange); introducing 
spin echoes (light green); and rescaling (dark green). The same is seen in the plots for the imaginary part and for $S(\omega)$, where results successively including these three classes of techniques are superposed. (Error mitigation acronyms: spin echoes (SE), Floquet calibration (F), readout error mitigation (RO), post-selection (PS)).
\label{fig:kh6}
}
\end{figure*}

\noindent \textbf{Ruthenium trichloride, $\alpha$-RuCl$_3$.} Fig.~\ref{fig:kh6} shows simulations for 6-site (7-qubit) and 10-site (11-qubit) models of $\alpha$-RuCl$_3$. Including the ancilla qubit, the latter system corresponds to the largest simulations we performed.

We first start with 6-site simulations of the heat capacity, 
obtained by numerically differentiating the finite-temperature energies.
The energy and heat capacity are shown in Fig.~\ref{fig:kh6}A. 
One of the main features of the proximate spin-liquid behaviour of $\alpha$-RuCl$_3$ is a two-peak structure in the heat capacity. 
While the energy has a significant error at lower temperatures, the two-peak structure of the heat capacity is visible 
at around $T=1$ and $T$ = 6-7, although it
is extremely noisy, as the numerical derivative amplifies the noise.

\begin{figure*}[!htbp]
\centering{
\includegraphics[width=0.95\textwidth]{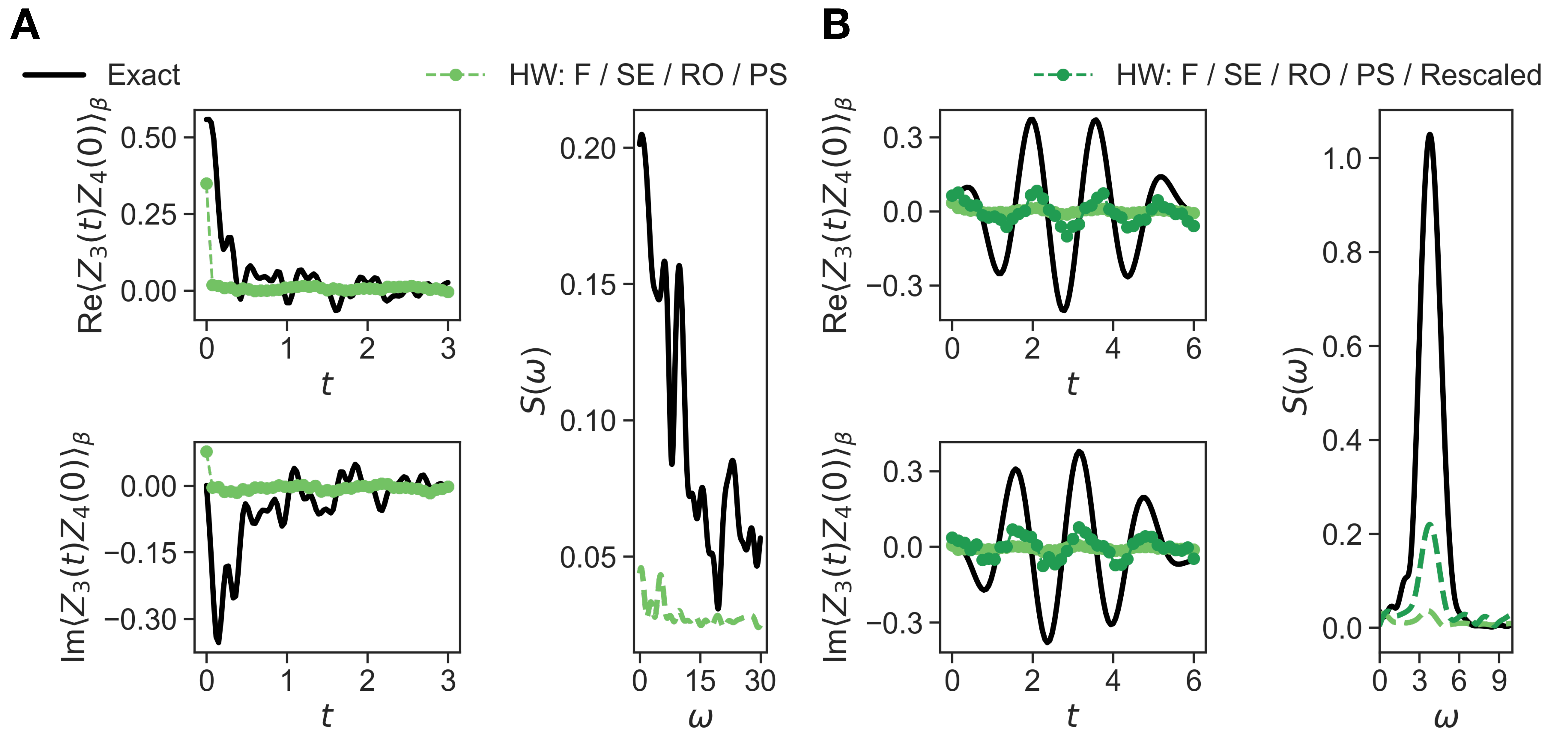}}
\caption{(A) Dynamical correlation function $\langle$$Z_3(t)Z_4(0)$$\rangle_\beta$ and its transform $S(\omega)$ at $\beta$ = 1 for $\alpha$-RuCl$_3$, 10-site model. Even after all error mitigation strategies (light green) the data in this case is too degraded to rescale.
(B) Dynamical correlation function $\langle$$Z_3(t)Z_4(0)$$\rangle_\beta$ and its transform $S(\omega)$ at $\beta$ = 1 for strongly anisotropic Kitaev-Heisenberg model parameters: $J$ = 0.4, $K_x$ = -8, $K_y$ = $K_z$ = $\frac{K_x}{6}$.
Reasonable frequencies are obtained after rescaling (dark green).
(Error mitigation acronyms: spin echoes (SE), Floquet calibration (F), readout error mitigation (RO), post-selection (PS)). }
\label{fig:kh10}
\end{figure*}

Fig.~\ref{fig:kh6}B and Fig.~\ref{fig:kh10}A show the dynamical correlation functions for the $\alpha$-RuCl$_3$ 6- and 10-site models (64 and 310 two-qubit gates, respectively), as well as a comparison to a second anisotropic parameter set (with smaller YY and ZZ terms) in Fig.~\ref{fig:kh10}B. Similarly to in the FeS clusters, the frequency structure is reasonably well preserved, with error mitigation and post-processing restoring the amplitudes for the 6-site model. However, in the 10-site $\alpha$-RuCl$_3$ model the data is too degraded to obtain any reasonable physical spectra. Fig.~\ref{fig:kh10}B shows the sensitivity of the quality of simulation to the choice of model, as the more anisotropic Kitaev-Heiseberg simulations are of significantly higher quality.

\section*{Discussion}
In the current work we discussed the quantum simulation of two representative real-world problems: the Fe-S clusters of nitrogenase, including the P-cluster and FeMo-co, a problem of interest in correlated quantum chemistry, and the proximate spin-liquid, $\alpha$-RuCl$_3$, a correlated material. These systems were not selected for their suitability for implementation on the Sycamore platform. Some reasonable results were obtained for simplified models of these problems. In the Fe-S clusters and the smaller $\alpha$-RuCl$_3$ instance, qualitatively correct features in the spin structure, excited-state spectrum, and heat capacity could be obtained. However, to achieve this, the implemented circuits needed to be obtained with the help of classical recompilation, and the data required significant processing, including using data from exact classical simulations of related tractable problems. Unfortunately, these steps raise questions with regards to effectively simulating more classically difficult systems.

The main limitation in the experiments was the two-qubit gate count, rather than the number of qubits. Simulations with up to 100 two-qubit gates, such as the larger Fe-S cluster simulations and the 6-site $\alpha$-RuCl$_3$ simulations, could be carried out with some confidence on the hardware. However, our largest simulations for $\alpha$-RuCl$_3$ which used 11 qubits, 310 two-qubit gates, and 782 single-qubit gates, were not successful. However, we could obtain meaningful simulation data with these quantum resources for Hamiltonian parameters that were tuned away from the $\alpha$-RuCl$_3$ regime, indicating the impact of tuning the problem for the characteristics of the hardware. The successfully deployed circuit resources were less than those used in some recent simulation experiments using a similar chip \cite{Google_FermiHubbard_arxiv_2020}. However, the discrepancy may
also be understood in terms of how well the experiment matches the
capabilities of the chip. For example, the simulations of the free fermion dynamics in ~\cite{Google_FermiHubbard_arxiv_2020} do not require microwave gates, which helps with Floquet calibration, and allows  more gates to be used. 

The questions of interest in realistic molecular and materials simulations are ones which require some degree of quantitative precision. Compared to the resources deployable to achieve random circuit supremacy on the same chip architecture, our representative simulation problems could use about 1/5 of the qubit and gate resources. If we adjusted our models to be more tuned to the hardware, it was possible to use more than 1/2 the gate resources of supremacy experiments, while retaining some level of physical accuracy. This provides an understanding of the relevance of artificial supremacy experiments to problems of physical simulation, and reflects the current status of quantum hardware and quantum simulation.

\section*{Data availability}
The data that support the findings of this study are available from the corresponding author upon reasonable request.

\section*{Code availability}
The code used to generate the numerical results presented in this paper can be made available upon reasonable request.

\section*{Acknowledgements}
\textbf{Funding:} R.N.T., R.H., and G.K.-L.C. were supported by the US Department of Energy, Office of Basic Energy Sciences, under Award No.~DE-SC0019374. S.-N.S., A.T.K.T, A.J.M were supported by the US NSF under Award No.~1839204. The quantum hardware used in this work was developed by the Google Quantum AI team. Data was collected via cloud access through Google’s Quantum Computing Service. The datasheet of the Weber device can be found at http://quantumai.google/hardware/datasheet/weber.pdf.

\section*{Author contributions}
R.N.T., R.H., and G.K.-L.C. conceptualized the project. 
R.N.T. and S.-N.S. designed and implemented the circuits with help from R.H.. R.N.T. executed the simulations and analyzed the results. R.N.T. and G.K.-L.C. wrote the paper. All authors discussed the results and contributed to the development of the manuscript.

\section*{Competing interests}
G.K.-L.C. is a part owner of QSimulate, Inc.

\clearpage
\bibliography{abbr,biblist}
\bibliographystyle{Science}

\clearpage
\setcounter{figure}{0}
\renewcommand{\thefigure}{S\arabic{figure}}
\setcounter{equation}{0}
\renewcommand{\theequation}{S\arabic{equation}}

\title{\textmd{Supplementary Information for}\\Simulating challenging correlated molecules and materials on the Sycamore quantum processor}

\author
{Ruslan N.~Tazhigulov,$^{1}$ Shi-Ning Sun,$^{2}$ Reza Haghshenas$^{1}$, Huanchen Zhai,$^{1}$ \\
Adrian~T.~K.~Tan,$^{2}$ Nicholas C.~Rubin,$^{3}$ Ryan Babbush,$^{3}$ Austin~J.~Minnich,$^{2}$\\
Garnet~Kin-Lic Chan$^{1}$\\
{\normalsize $^{1}$Division of Chemistry and Chemical Engineering, California Institute}\\ \normalsize{of Technology, Pasadena, California 91125, USA}\\
\normalsize{$^{2}$Division of Engineering and Applied Science, California Institute of}\\ \normalsize{Technology, Pasadena, California 91125, USA}\\
\normalsize{$^{3}$Google Quantum AI, 340 Main Street, Venice, California 90291, USA}
}
\date{}


\baselineskip24pt

\maketitle

\section{Spectrum of $S=5/2$ Fe-S models versus $S=1/2$ Fe-S models}

\begin{figure}[!htbp]
\centering
  \begin{tikzpicture}[
    ax/.style={black, line width=1.5pt},
    axx/.style={black!20, line width=1.5pt},
    lv/.style={sing,line width=1pt},
    lu/.style={trip,line width=1pt},
    lw/.style={quin, line width=1pt},
    lx/.style={sept, line width=1pt},
    ly/.style={none, line width=1pt},
    cn/.style={dotted,black!40, line width=1pt},
    xn/.style={align=center}]
    \definecolor{mxc}{RGB}{120,160,160};
    \draw[axx] (0.25, 0) -- (2.45, 0);
    \draw[axx] (3.2, 0) -- (5.5, 0);
    \draw[ax,->] (0, 0) -- (0, 6);
    \node[rotate=90] at (-0.75,2.5) {Excitation Energy};
    \foreach \x in {0,1,2,3} {
        \draw[ax] (-0.15, {\x * 1.5}) -- (0.02, {\x * 1.5});
        \node[left] at (-0.15, {\x * 1.5}) {\x}; }
    \draw[lv] ({1.25},{0.000 * 1.5}) -- ({1.2 + 0.25},{0.000 * 1.5});
    \draw[lv] ({1.25},{0.340 * 1.5}) -- ({1.2 + 0.25},{0.340 * 1.5});
    \foreach[count=\i] \y in {0,1}
    \draw[lu] ({0.75 + \i * 0.25 + 0.125},{1.170 * 1.5})
        -- ({0.95 + \i * 0.25 + 0.125},{1.170 * 1.5});
    \draw[lu] ({1.25},{1.340 * 1.5}) -- ({1.2 + 0.25},{1.340 * 1.5});
    \draw[lw] ({1.25},{3.340 * 1.5}) -- ({1.2 + 0.25},{3.340 * 1.5});
    \draw[lv] ({4.25},{0.000 * 1.5}) -- ({4.2 + 0.25},{0.000 * 1.5});
    \draw[lv] ({4.25},{0.340 * 1.5}) -- ({4.2 + 0.25},{0.340 * 1.5});
    \draw[lv] ({4.25},{1.020 * 1.5}) -- ({4.2 + 0.25},{1.020 * 1.5});
    \draw[lv] ({4.25},{2.040 * 1.5}) -- ({4.2 + 0.25},{2.040 * 1.5});
    \draw[lv] ({4.25},{3.400 * 1.5}) -- ({4.2 + 0.25},{3.400 * 1.5});
    \draw[lv] ({4.25},{3.720 * 1.5}) -- ({4.2 + 0.25},{3.720 * 1.5});
    \foreach[count=\i] \y in {0,1}
    \draw[lu] ({3.75 + \i * 0.25 + 0.125},{1.170 * 1.5})
        -- ({3.95 + \i * 0.25 + 0.125},{1.170 * 1.5});
    \draw[lu] ({4.25},{1.340 * 1.5}) -- ({4.2 + 0.25},{1.340 * 1.5});
    \foreach[count=\i] \y in {0,1}
    \draw[lu] ({3.75 + \i * 0.25 + 0.125},{1.680 * 1.5})
        -- ({3.95 + \i * 0.25 + 0.125},{1.680 * 1.5});
    \draw[lu] ({4.25},{2.020 * 1.5}) -- ({4.2 + 0.25},{2.020 * 1.5});
    \foreach[count=\i] \y in {0,1}
    \draw[lu] ({3.75 + \i * 0.25 + 0.125},{2.530 * 1.5})
        -- ({3.95 + \i * 0.25 + 0.125},{2.530 * 1.5});
    \draw[lu] ({4.25},{3.040 * 1.5}) -- ({4.2 + 0.25},{3.040 * 1.5});
    \draw[lw] ({4.25},{3.340 * 1.5}) -- ({4.2 + 0.25},{3.340 * 1.5});
    \foreach[count=\i] \y in {0,1}
    \draw[lw] ({3.75 + \i * 0.25 + 0.125},{3.510 * 1.5})
        -- ({3.95 + \i * 0.25 + 0.125},{3.510 * 1.5});
    \foreach[count=\i] \y in {0,1}
    \draw[lw] ({3.75 + \i * 0.25 + 0.125},{3.680 * 1.5})
        -- ({3.95 + \i * 0.25 + 0.125},{3.680 * 1.5});
    \node[xn] at(1.25, -0.45) {$S=\frac{1}{2}\mathrm{\ model}$};
    \node[xn] at(4.25, -0.45) {$S=\frac{5}{2}\mathrm{\ model}$};
    \draw[cn] (1.65, {0.340 * 1.5}) -- (4.05, {0.340 * 1.5});
    \draw[cn] (1.775, {1.170 * 1.5}) -- ({4.05-0.125}, {1.170 * 1.5});
    \draw[cn] (1.65, {1.340 * 1.5}) -- (4.05, {1.340 * 1.5});
    \draw[cn] (1.65, {3.340 * 1.5}) -- (4.05, {3.340 * 1.5});
    \draw[lv] (5.5, {10.5 / 2}) -- (5.8, {10.5 / 2})
        node[right] {singlet};
    \draw[lu] (5.5, {10.5 / 2 - 0.4}) -- (5.8, {10.5 / 2 - 0.4})
        node[right] {triplet};
    \draw[lw] (5.5, {10.5 / 2 - 0.8}) -- (5.8, {10.5 / 2 - 0.8})
        node[right] {quintet};
  \end{tikzpicture}
  \caption{The spectra of the \( S = 1/2 \) and \( S = 5/2\) Heisenberg models for the [4Fe-4S] cluster.}
  \label{fig:spec1}
\end{figure}
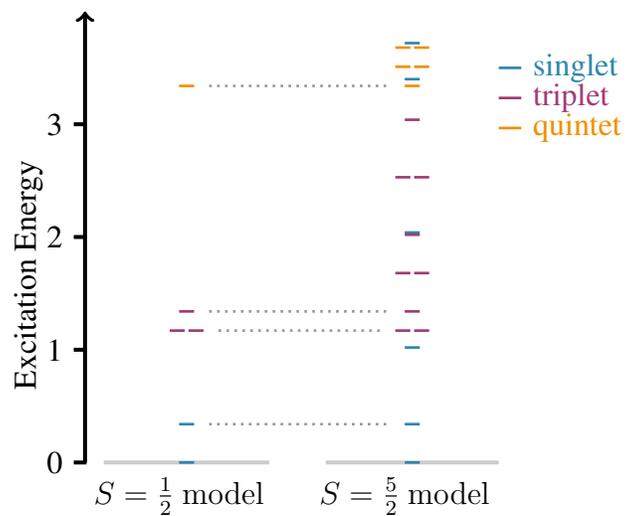

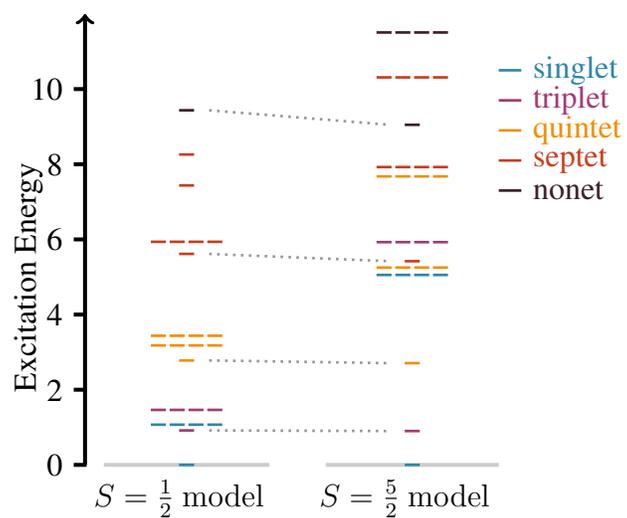
\begin{figure}[!htbp]
\centering
  \begin{tikzpicture}[
    ax/.style={black, line width=1.5pt},
    axx/.style={black!20, line width=1.5pt},
    lv/.style={sing,line width=1pt},
    lu/.style={trip,line width=1pt},
    lw/.style={quin, line width=1pt},
    lx/.style={sept, line width=1pt},
    ly/.style={none, line width=1pt},
    cn/.style={dotted,black!40, line width=1pt},
    xn/.style={align=center}]
    \definecolor{mxc}{RGB}{120,160,160};
    \draw[axx] (0.25, 0) -- (2.45, 0);
    \draw[axx] (3.2, 0) -- (5.5, 0);
    \draw[ax,->] (0, 0) -- (0, 6);
    \node[rotate=90] at (-0.75,2.5) {Excitation Energy};
    \foreach \x in {0,2,4,6,8,10} {
        \draw[ax] (-0.15, {\x / 2.0}) -- (0.02, {\x / 2.0});
        \node[left] at (-0.15, {\x / 2.0}) {\x}; }
    \draw[lv] ({1.25},{0.000 / 2}) -- ({1.2 + 0.25},{0.000 / 2});
    \draw[lu] ({1.25},{0.917 / 2})
        -- ({1.2 + 0.25},{0.917 / 2});
    \foreach[count=\i] \y in {0,1,2,3}
    \draw[lv] ({0.75 + \i * 0.25 - 0.125},{1.070 / 2})
        -- ({0.95 + \i * 0.25 - 0.125},{1.070 / 2});
    \foreach[count=\i] \y in {0,1,2,3}
    \draw[lu] ({0.75 + \i * 0.25 - 0.125},{1.463 / 2})
        -- ({0.95 + \i * 0.25 - 0.125},{1.463 / 2});
    \draw[lw] ({1.25},{2.779 / 2}) -- ({1.2 + 0.25},{2.779 / 2});
    \foreach[count=\i] \y in {0,1,2,3}
    \draw[lw] ({0.75 + \i * 0.25 - 0.125},{3.179 / 2})
        -- ({0.95 + \i * 0.25 - 0.125},{3.179 / 2});
    \foreach[count=\i] \y in {0,1,2,3}
    \draw[lw] ({0.75 + \i * 0.25 - 0.125},{3.436 / 2})
        -- ({0.95 + \i * 0.25 - 0.125},{3.436 / 2});
    \draw[lx] ({1.25},{5.614 / 2}) -- ({1.2 + 0.25},{5.614 / 2});
    \foreach[count=\i] \y in {0,1,2,3}
    \draw[lx] ({0.75 + \i * 0.25 - 0.125},{5.936 / 2})
        -- ({0.95 + \i * 0.25 - 0.125},{5.936 / 2});
    \draw[lx] ({1.25},{7.436 / 2}) -- ({1.2 + 0.25},{7.436 / 2});
    \draw[lx] ({1.25},{8.259 / 2}) -- ({1.2 + 0.25},{8.259 / 2});
    \draw[ly] ({1.25},{9.436 / 2}) -- ({1.2 + 0.25},{9.436 / 2});
    \draw[lv] ({4.25},{0.000 / 2}) -- ({4.2 + 0.25},{0.000 / 2});
    \draw[lu] ({4.25},{0.901 / 2}) -- ({4.2 + 0.25},{0.901 / 2});
    \foreach[count=\i] \y in {0,1,2,3}
    \draw[lv] ({3.75 + \i * 0.25 - 0.125},{5.054 / 2})
        -- ({3.95 + \i * 0.25 - 0.125},{5.054 / 2});
    \foreach[count=\i] \y in {0,1,2,3}
    \draw[lu] ({3.75 + \i * 0.25 - 0.125},{5.927 / 2})
        -- ({3.95 + \i * 0.25 - 0.125},{5.927 / 2});
    \draw[lw] ({4.25},{2.707 / 2}) -- ({4.2 + 0.25},{2.707 / 2});
    \foreach[count=\i] \y in {0,1,2,3}
    \draw[lw] ({3.75 + \i * 0.25 - 0.125},{5.250 / 2})
        -- ({3.95 + \i * 0.25 - 0.125},{5.250 / 2});
    \foreach[count=\i] \y in {0,1,2,3}
    \draw[lw] ({3.75 + \i * 0.25 - 0.125},{7.676 / 2})
        -- ({3.95 + \i * 0.25 - 0.125},{7.676 / 2});
    \draw[lx] ({4.25},{5.420 / 2}) -- ({4.2 + 0.25},{5.420 / 2});
    \foreach[count=\i] \y in {0,1,2,3}
    \draw[lx] ({3.75 + \i * 0.25 - 0.125},{7.925 / 2})
        -- ({3.95 + \i * 0.25 - 0.125},{7.925 / 2});
    \foreach[count=\i] \y in {0,1,2,3}
    \draw[lx] ({3.75 + \i * 0.25 - 0.125},{10.311 / 2})
        -- ({3.95 + \i * 0.25 - 0.125},{10.311 / 2});
    \draw[ly] ({4.25},{9.049 / 2}) -- ({4.2 + 0.25},{9.049 / 2});
    \foreach[count=\i] \y in {0,1,2,3}
    \draw[ly] ({3.75 + \i * 0.25 - 0.125},{11.506 / 2})
        -- ({3.95 + \i * 0.25 - 0.125},{11.506 / 2});
    \node[xn] at(1.25, -0.45) {$S=\frac{1}{2}\mathrm{\ model}$};
    \node[xn] at(4.25, -0.45) {$S=\frac{5}{2}\mathrm{\ model}$};
    \draw[cn] (1.65, {0.917/ 2}) -- (4.05, {0.901 / 2});
    \draw[cn] (1.65, {2.779/ 2}) -- (4.05, {2.707 / 2});
    \draw[cn] (1.65, {5.614/ 2}) -- (4.05, {5.420 / 2});
    \draw[cn] (1.65, {9.436/ 2}) -- (4.05, {9.049 / 2});
    \draw[lv] (5.5, {10.5 / 2}) -- (5.8, {10.5 / 2})
        node[right] {singlet};
    \draw[lu] (5.5, {10.5 / 2 - 0.4}) -- (5.8, {10.5 / 2 - 0.4})
        node[right] {triplet};
    \draw[lw] (5.5, {10.5 / 2 - 0.8}) -- (5.8, {10.5 / 2 - 0.8})
        node[right] {quintet};
    \draw[lx] (5.5, {10.5 / 2 - 1.2}) -- (5.8, {10.5 / 2 - 1.2})
        node[right] {septet};
    \draw[ly] (5.5, {10.5 / 2 - 1.6}) -- (5.8, {10.5 / 2 - 1.6})
        node[right] {nonet};
  \end{tikzpicture}
  \caption{The spectra of the \( S = 1/2 \) and \( S = 5/2\) Heisenberg models for the P/FeMo-co cluster.}
  \label{fig:spec2}
\end{figure}

As mentioned in the main text, the $S=1/2$ model Fe-S spectrum has some features of the more realistic $S=5/2$ model spectrum. This is illustrated in Figures~\ref{fig:spec1}, \ref{fig:spec2}.

\section{Depolarizing noise in the Fe-S simulation}

\begin{figure}
\centering
\includegraphics[width=4in]{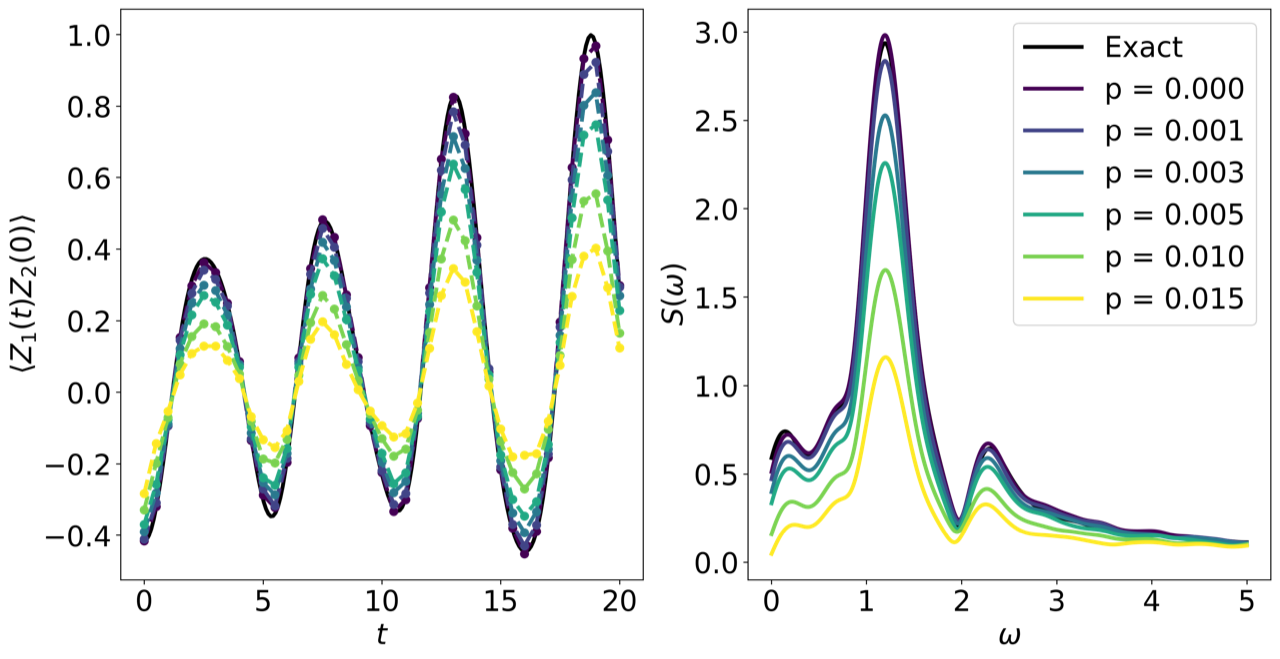}
\caption{Noisy emulation of the [4Fe-4S] model spectrum.
\label{fig:emul}}
\end{figure}

We model depolarizing noise in classical emulation by inserting symmetric depolarizing noise after each circuit moment:
\begin{align}
\rho \to (1 - 3p) \rho + p (X \rho X + Y \rho Y + Z \rho Z)
\end{align}
The result of such a noisy emulation is shown in Fig.~\ref{fig:emul}. Comparing to the data in Fig.~4B, we see that the hardware result without error mitigation can be reproduced with a depolarizing noise value of $p=0.005$-0.010.

\end{document}